\documentclass[showpacs,twocolumn,amsmath,amssymb]{revtex4}

\usepackage{graphicx}
\usepackage{dcolumn}
\usepackage{bm}
\usepackage{hyperref}
\usepackage{latexsym}

\begin{document}

\title{\bf Nanopercolation}

\author{J. S. Andrade Jr.$^{1}$, D. L. Azevedo$^{1,2}$, R. Correa Filho$^{1}$,
and R. N. Costa Filho$^{1}$}
\affiliation{$^1$Departamento de F\'{\i}sica, Universidade Federal
do Cear\'a, 60451-970 Fortaleza, Cear\'a, Brazil.\\
$^2$Departamento de F\'{\i}sica, Universidade Federal do Maranh\~{a}o,\\
65080-040 S\~ao Luis, Maranh\~ao, Brazil\\}

\date{\today}

\begin{abstract}
We investigate through direct molecular mechanics calculations the
geometrical properties of hydrocarbon mantles subjected to percolation
disorder. We show that the structures of mantles generated at the critical
percolation point have a fractal dimension $d_{f} \approx 2.5$. In addition, the
solvent access surface $A_{s}$ and volume $V_{s}$ of these molecules follow
power-law behavior, $A_{s} \sim L^{\alpha_A}$ and $V_{s} \sim L^{\alpha_V}$,
where $L$ is the system size, and with both critical exponents $\alpha_A$ and
$\alpha_V$ being significantly dependent on the radius of the
accessing probing molecule, $r_{p}$. Our results from extensive
simulations with two distinct microscopic topologies (i.e., square and
honeycomb) indicate the consistency of the statistical analysis and
confirm the self-similar characteristic of the percolating
hydrocarbons. Due to their highly branched topology, some of the
potential applications for this new class of disordered molecules
include drug delivery, catalysis, and supramolecular structures.
\end{abstract}


\maketitle

In recent years there has been a great scientific and technological
advance in the design and development of new materials. The use of
molecular modelling techniques is certainly responsible for the
significant contribution to this activities of most theoretical and
computational studies \cite{Leach01}. New improved catalysts, original
reactive processing and the chemistry and physics of polymers are just
a few among several examples representing the multitude of research
fields that have been successfully approached employing molecular
modelling tools. For instance, an important research work with
potential applications in the field of polymer science refers to the
real possibility of manipulating and controlling the properties of
complex macromolecules through adequate polymerization routes and
specific chemical processing strategies.

An inspiring example where molecular modelling plays an important role
in polymer science is the design and characterization of {\it
 functional}
polymers. By definition, the term functional refers to the property of
some molecules that contain reactive end-groups to preserve the
integrity of their molecular backbone while participating in chemical
reactions. Because {\it dendrimers} are highly branched
macromolecules and, therefore, have a large number of end-groups, they
represent a natural class of candidates for substrates of functional
polymers. Among many possible applications for dendrimers, we mention
their potential use for drug delivery \cite{Esfand01}, and the ability
to act as catalysts supports \cite{Brunner95} and supramolecular structures
\cite{Emrick99}. In a recent theoretical study, the molecular topology
and surface accessibility of dendrimers have been investigated through
extensive molecular mechanics and molecular dynamics simulations
\cite{Pricl03}. Accordingly, it has been shown that all dendrimer
families exhibit non-spherical and shape-persistent structures, with a
molecular fractal dimension close to $2.5$ and an invariant surface
fractal dimension approximately equal to $2$, regardless the number
of generations (branches) composing the molecular geometry.

In the present letter, we propose a new class of macromolecules that
are also highly branched, but possess intrinsic disorder in their
compositional and structural conformation. Here we use percolation as
a model for disorder. Due to its simplicity, the percolation
theory \cite{Stauffer94,Sahimi94} has been extensively applied to
represent disordered materials in several research fields of
scientific and technological relevance, such as condensed matter
physics \cite{Nakayama03}, flow through porous media \cite{Sahimi95},
and heterogeneous chemistry\cite{Avnir89,Havlin96,Andrade97a,Costa03},
among others. More precisely, percolation is a purely geometric model
in the sense that we have to populate all lattice sites (or bonds)
with a prescribed probability $p$. For small values of $p$, only
finite clusters are present. By increasing $p$, one can find a
threshold or a critical value $p_c$ for which an ``infinitely''
connected object, the {\it spanning cluster}, is generated. In fact,
the critical geometry of the spanning cluster\cite{Stauffer94,Havlin96}
is an example of random fractal that has been frequently used as a
representation for real disordered systems. The aim here is to
investigate through molecular mechanics simulations the geometrical
properties of hydrocarbon molecules generated from carbon mantles
subjected to site percolation disorder.

Starting from a square or an hexagonal lattice of size $L$, we extract
the spanning cluster and assign to each site a carbon atom that is
connected by single bonds to its nearest neighbors. The valence of
each carbon in the cluster is then adjusted to 4 by adding the
necessary hydrogen atoms, so that, in average, a hydrocarbon with
fractional stoichiometry is produced, namely, {\bf C}$_x${\bf H}$_y$.
The bond length in the initial configuration of this molecule is set
to an arbitrary guess value (e.g., $1$\AA) and a molecular mechanics
method is applied to optimize the molecular geometry \cite{Leach01}.
Molecular mechanical force fields use the equations of classical
mechanics to describe the potential energy surfaces and physical
properties of molecules.  Under this framework, a molecule is viewed
as a collection of atoms that interact with each other by simple
analytical functions. Here we apply the standard {\bf MM+} force field
\cite{Allinger77} to obtain the optimized geometry for each
realization of the percolating hydrocarbon. For comparison, a limited
number of computational simulations have also been performed with a
more realistic approach, namely, the {\bf PM3} semi-empirical and
self-consistent-field molecular orbital method \cite{Stewart89}. For
all practical purposes, the relative differences found between the
geometrical properties calculated here resulting from the {\bf PM3}
technique and those obtained using the {\bf MM+} molecular mechanics
are always smaller than $3\%$.

In Fig.~1a we show the the original square lattice of carbon sites
(planar) and the optimized geometries computed for a carbon mantle
that is fully occupied ($p=1$). When submitted to the optimization
procedure, the molecule bends spontaneously to assume the shape of a
corrugated tile. In the case of an hexagonal lattice, the resulting
molecular structure remains planar. The structures shown in Figs.~1b
and 1c correspond to typical configurations of hydrocarbons mantles
generated from square and hexagonal lattices, respectively, but now
with the probability $p$ set right above the critical percolation
point $p_c$.

\begin{figure}
\resizebox{0.5\textwidth}{!}{\includegraphics{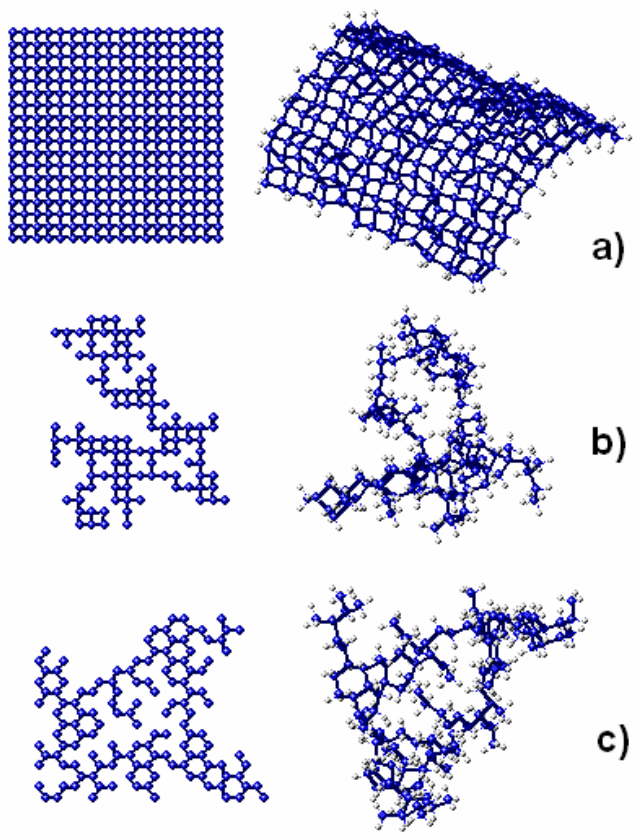}}
\caption{(a) Hydrocarbon mantle of size $L=18$ generated from a fully
occupied square lattice ($p=1$). The hydrocarbon chains on the left
are the initial two-dimensional structures used in the optimization
procedure, while those on the right are the corresponding three-dimensional
topologies resulting from the molecular mechanics simulations. (b) The same
as in (a), but for a value of $p$ right above the critical percolation point
$p_c$. (c) The same as in (b), but for an hexagonal lattice topology.}
\end{figure}

We performed simulations for square and hexagonal lattices generated
at the critical point ($p_c \approx 0.593$ and $0.696$, respectively),
and with linear sizes ranging from $L=10$ to $32$. In the case of the
square lattice, for each linear size $L=12$, $15$, $18$, $21$, $24$,
$27$, and $30$, we used $300$, $300$, $150$, $150$, $80$, $80$, and
$50$ realizations, respectively. For the hexagonal geometry, the
calculations have been carried out with $300$, $150$, $150$, $80$,
$80$, and $50$ realizations for $L=10$, $14$, $18$, $22$, $26$, and
$30$, respectively. For each realization, once the convergence for the
minimum energy structure is numerically attained, we compute a set of
relevant geometrical properties that are then averaged over all
samples of same size $L$.  For instance, the radius of gyration
$R_{g}$ represents a meaningful measure to characterize objects of
complicated geometry \cite{Stauffer94}. It is defined here as
$R_{g}^{2}\equiv\sum_{i=1}^{N_{a}}|{\bf{{r}_{i}}-\bf{{r}_{0}}|}^{2}/N_{a}$,
where $N_{a}$ is the number of atoms,
$\bf{{r}_{0}}\equiv\sum_{i=1}^{N_{a}}\bf{{r}_{i}}$, and $\bf{r}_{i}$ is the
position of each atom $i$ in the molecule, regardless of its type, i.e.,
hydrogen or carbon. The results shown in Fig.~2 clearly indicate that the radius
of gyration displays a typical power-law behavior for both networks topologies,
\begin{equation}
R_{g} \sim M^{1/d_{f}}~,
\end{equation}
where $M$ is the average molecular weight and $d_{f}$ is the fractal
dimension of the disordered molecules. The best fits to the data yield
exponents that are statistically similar, namely, $0.40
\pm 0.03$ and $0.38 \pm 0.02$ corresponding to the fractal dimensions
$d_{f}=2.50$ and $2.63$, for the square and hexagonal lattices,
respectively. Considering the small sizes of the lattices employed in
the simulations due to computational limitations, the agreement
between these two exponents suggests that the complex geometry of
percolating hydrocarbons may constitute a single class of
universality.

\begin{figure}
\includegraphics[width=8.0cm]{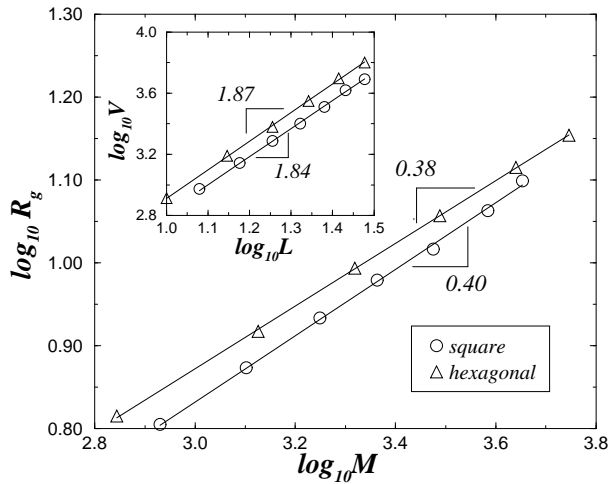}
\caption{Logarithmic plot showing the dependence of the radius of
 gyration
$R_{g}$ on the average molecular weight $M$ of percolating
hydrocarbons generated from square (circles) and hexagonal (triangles)
lattices. The straight lines are the least-square fits to the data
with the numbers indicating the slopes, $0.40 \pm 0.03$ (circles) and
$0.38 \pm 0.02$ (triangles). The inset shows the log-log plot of the
van der Waals volume against the linear dimension of the system for
square (circles) and hexagonal (triangles) lattices.}
\end{figure}

At the microscopic scale it is important to determine the volume and
surface of a molecule. Here we use the concept of van der Waals volume
(surface) which is the sum of the volume (surface) of each atom
composing a molecule, calculated as a sphere with the corresponding
van der Waals radius. By definition, the van der Waals radii are
simply the radii of the spheres connecting two non-bonded atoms
\cite{Leach01}. The inset of Fig.~2 shows the log-log plot of the van
der Walls volume $V_{W}$ against the size $L$ of the system for both
square and hexagonal lattices. From the least-square fits to the two
data sets we observe that the results for both lattice topologies can
be well described by the scaling relation, $V_{W} \sim L^{d_{W}}$,
where $d_{W}$ is the critical exponent. If we consider that each
carbon gives approximately the same contribution to the overall volume
of the molecule, the values $d_{W}=1.84 \pm 0.03$ and $1.87 \pm 0.02$
obtained for the square and hexagonal lattice, respectively, can be
readily compared with the fractal dimension of the spanning cluster at
the critical percolation point, which is $\approx 1.89$ in two
dimensions \cite{Stauffer94}. The small discrepancies found here are
due to finite-size effects as well as the presence of the hydrogens
atoms necessary to balance the valence of the carbon sites.

As already mentioned, the investigation of molecular volume and
surface is essential for the development of many chemical processes of
scientific and technological relevance. Indeed, the molecular
architecture can play a major role in determining the potential
application and efficiency of functional macromolecules
\cite{Pricl03,Wang04}. In relation to that, the concepts of solvent
accessible surface $A_{s}$ and volume $V_{s}$ \cite{Connolly85}
have been extensively used to characterize and compare the geometrical
aspects of the interaction between a molecule and different types of
solvent. As depicted in Fig.~3, the $A_{s}$ is obtained by rolling a
spherical probe of diameter $2r_{p}$ corresponding to the size of the
solvent (e.g., water or ethanol) on the van der Waals surface
($A_{W}$) of the molecule. The $V_{s}$ is the volume for which the
boundary is the $A_{s}$. Since macromolecules like the percolating
hydrocarbons studied here contain small gaps, pockets and clefts which
are sometimes too small to be penetrated even by a solvent molecule
like water, the $A_{s}$ becomes gradually smoother as the size of the
solvent used for probing increases. In the limiting case where $r_{p}$
is set to zero, we recover the $A_{W}$ of the molecule.

\begin{figure}
\resizebox{0.5\textwidth}{!}{\includegraphics{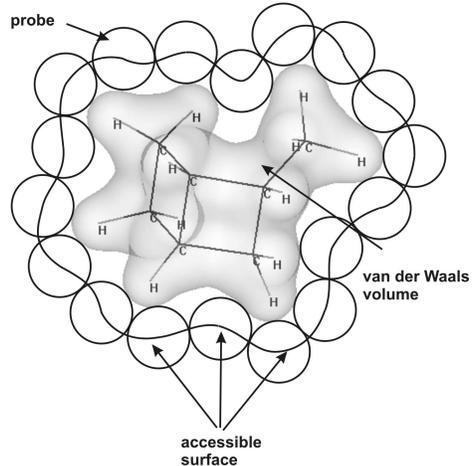}}
\caption{The solvent accessible surface and volume of each realization
of a percolating hydrocarbon are obtained by rolling a spherical probe
of radius $r_{p}$ on its van der Waals surface. For each lattice
topology (square or hexagonal) and a given value of the linear
dimension $L$, the average surface $A_{s}$ and volume $V_{s}$ are
calculated from several realizations of the disordered molecules.}
\end{figure}

\begin{figure}
\begin{center}
\includegraphics[width=8.0cm]{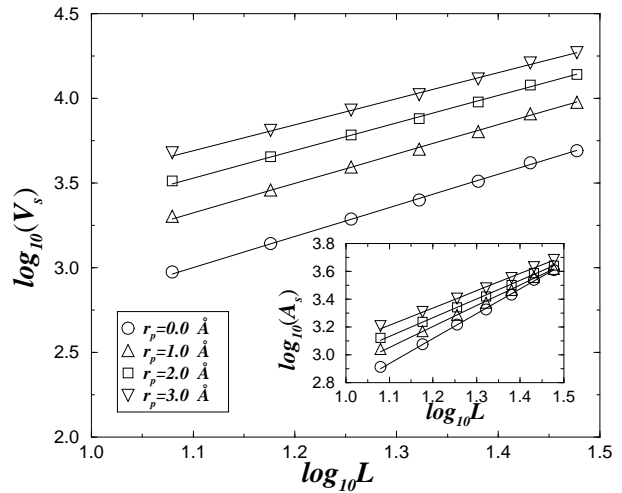}
\caption{Log-log plot of the variation of the average accessible
 volume
$V_{s}$ with the linear dimension $L$ for four different values of the
solvent probe radius $r_{p}$. The inset shows the same variation, but
for the average accessible surface $A_{s}$. The straight lines are the
best fits to the data sets of the scaling relations $A_{s} \propto
L^{\alpha_A}$ and $V_{s} \propto L^{\alpha_V}$, with the slopes
corresponding to the exponents $\alpha_A$ and $\alpha_V$ . These
results are for percolating hydrocarbons generated from a square
lattice topology.}
\end{center}
\end{figure}

In Fig.~4 we show the log-log plot of the average $V_{s}$ against the
linear size of percolating hydrocarbons generated from square
lattices, and calculated for four different values of $r_{p}$. These
results provide clear evidence that the $V_{s}$ of such disordered
molecules increases as a power-law,
\begin{equation}
V_{s} \sim L^{\alpha_{V}}~,
\end{equation}
with an exponent $\alpha_{V}$ that significantly depends on the size
$r_{p}$ of the probing molecule. The inset of Fig.~4 shows that the
$A_{s}$ also follows a power-law, $A_{s} \sim L^{\alpha_{A}}$, with an
exponent $\alpha_{A}$ that is, like in the $V_{s}$ case, nonuniversal
with respect to the solvent radius. We find an entirely similar
behavior for both $V_{s}$ and $A_{s}$ obtained from molecular
mechanics simulations performed with critical hexagonal lattices.

Figure~5 shows the dependence of $\alpha_{V}$ and $\alpha_{A}$ on the
radius $r_{p}$ for the case of the square topology. Starting from
values that are close to the fractal dimension of the spanning cluster
in two-dimensions, both exponents decrease monotonically with $r_{p}$,
reflecting the gradual smoothness of the $A_{s}$ as the size of the
solvent molecule increases. In all cases, we find a maximum relative
deviation of $5\%$ between scaling exponents of square and hexagonal
lattices computed at the same value of $r_{p}$. As expected, these
systems belong to the same universality class, despite the difference
in the details of their microscopic geometry as well as the fact that
a larger contribution of hydrogen atoms is needed to balance the
critical hexagonal molecules, in comparison to their square
counterparts of same size. These results indicate that any type of
physical or chemical interaction between a percolating hydrocarbon and
a given molecule can be dramatically dependent on the molecule
size. For instance, even if uniformly functionalized, these highly
structured hydrocarbons may display enormous discrepancies in
reactivity if the reagents have slightly different sizes.

\begin{figure}
\includegraphics[width=8.0cm]{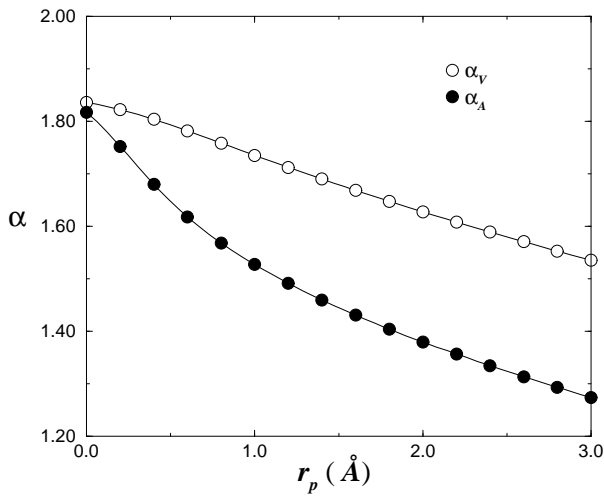}
\caption{Dependence of the scaling exponents $\alpha_A$ and $\alpha_V$
on the solvent probe radius $r_{p}$ for percolating hydrocarbons
extracted from a square lattice.}
\end{figure}

In conclusion we have studied the average geometrical features of
molecules generated from hydrocarbon mantles subjected to percolation
disorder. Our results show that these nanostructures are self-similar
and display a rich variety of scaling behavior in their solvent
accessible volume and surface properties. We expect these disordered
molecules to be experimentally feasible. As an important step in this
direction, ultrathin sheet-like carbon nanostructures, also called
carbon nanosheets, have been recently synthesized under high
controllability by a plasma enhanced process of chemical vapor
deposition \cite{Wang04}. Among other characteristics, these molecules
have a very high surface-to-volume ratio and sharp edges which are
attractive for fuel cells and micro-electronic technologies. Under a
different perspective, Andrade {\it et al.} \cite{Andrade97b} proposed
a growth mechanism for branched polymers where self-organization leads
the system spontaneously to a percolation-like critical state that is
entirely similar to the disordered molecules investigated here. In
this study, a conceivable experimental scheme is suggested for the
implementation of the method that could lead to the generation of
percolation hydrocarbons.

We thank the Brazilian agencies CNPq, CAPES and FUNCAP for financial
 support.

\end{document}